\documentclass[12pt]{article}
\usepackage{amsmath}
\usepackage{amsfonts}
\usepackage{amssymb}
\usepackage{amsthm}
\usepackage{graphicx,psfrag,epsf}
\usepackage{enumerate}
\usepackage{natbib}
\usepackage{hyperref}
\usepackage{url}
\usepackage[title]{appendix}
\usepackage[plain,noend]{algorithm2e}

\bibpunct{(}{)}{;}{a}{}{,}

\newtheorem{proposition}{Proposition}
\newtheorem{corollary}{Corollary}
\newtheorem{assumption}{Assumption}

\begin{document}

\def\spacingset#1{\renewcommand{\baselinestretch}%
{#1}\small\normalsize} \spacingset{1}


\title{On the overestimation of widely applicable Bayesian information criterion}
  \author{Toru Imai\footnote{imai.toru.7w@kyoto-u.ac.jp} \\
    Kyoto University}
  \date{}
  \maketitle

\bigskip
\begin{abstract}
A widely applicable Bayesian information criterion \citep{Watanabe2013} is applicable for both regular and singular models in the model selection problem.
This criterion tends to overestimate the log marginal likelihood.
We identify an overestimating term of a widely applicable Bayesian information criterion.
Adjustment of the term gives an asymptotically unbiased estimator of the leading two terms of asymptotic expansion of the log marginal likelihood.
In numerical experiments on regular and singular models, the adjustment resulted in smaller bias than the original criterion.
\end{abstract}

\noindent%
\vfill

\spacingset{1.45} 

\section{Introduction}
Evaluation on the log marginal likelihood is an important issue in the model selection problem, and a number of studies have been conducted (see, for example, \cite{Konishi2007}). \cite{Schwarz1978} proposed the Bayesian information criterion, BIC, which gives an approximation of the log marginal likelihood. However, BIC requires regularity conditions and therefore covers only regular models.

On the other hand, \cite{Watanabe2013} proposed the widely applicable Bayesian information criterion, WBIC, which can be applied to both regular and singular models. Unfortunately, WBIC tends to overestimate the log marginal likelihood in numerical experiments \citep{Friel2017}. In order to prove this overestimation, we need to identify three components: the $O_p(1)$ term of WBIC, the multiplicity of the real log canonical threshold, and the $O_p(1)$ term of the log marginal likelihood. However, it is challenging to identify the second and third components.

The aim of this paper is to identify the explicit constant order term of WBIC that causes an overestimation.

\section{Widely applicable Bayesian information criterion}

Let $X^n=(X_1,...,X_n)$ denote a sample of $n$ independent and identically distributed observations with each $X_i \in R^h$ drawn from a data generating distribution $q$. Let $M$ be a $d$-dimensional model with associated parameters $\theta \in \Omega \subset R^{d}$, where $\Omega$ is a parameter space.
Let $p(X^n \!\mid\! \theta, M)$ be the likelihood function and $\varphi(\theta \!\mid\! M)$ a prior distribution. 
The log marginal likelihood $\log L(M)$ for model $M$ is defined as
\begin{equation*}
\log L(M) := \log \int_{\Omega} p(X^n \!\mid\! \theta, M) \varphi(\theta \!\mid\! M) d\theta.
\end{equation*}

A statistical model is termed regular if the mapping from a model parameter to a probability distribution is one-to-one and if the Fisher information matrix is positive definite. Otherwise, a statistical model is called singular. In this paper, we assume that $p(X^n \!\mid\! \theta, M)$ is differentiable and that its first derivative function is not a constant. 

For any integrable function $f(\theta)$ and a non-negative real variable $t$, let $E_\theta^t \{ f(\theta) \}$ and $V_\theta^t \{ f(\theta) \}$ be defined as
\begin{eqnarray*}
E_\theta^t \{ f(\theta) \} &=&  \left\{ \int_{\Omega} p(X^n \!\mid\! \theta, M)^t \varphi (\theta \!\mid\! M) d\theta \right\}^{-1 }\int_{\Omega} f(\theta) p(X^n \!\mid\! \theta, M)^t \varphi (\theta \!\mid\! M) d\theta,  \\
V_\theta^t \{ f(\theta) \} &=& E_\theta^t \{ f(\theta)^2 \} - \Bigl[E_\theta^t \{f(\theta) \} \Bigr]^2,
\end{eqnarray*}
respectively. Here, $t$ is called an inverse temperature.

Let $F(t)$ be defined as
\begin{equation*}
F(t) := \log \int_\Omega p(X^n \!\mid\! \theta, M)^t \varphi(\theta \!\mid\! M) d\theta.
\end{equation*}
Then, $F(0)=0, F(1)=\log L(M)$ by definition, and a simple calculation gives
\begin{eqnarray*}
\frac{d}{dt} F(t) &=& E_\theta^t \{ \log p(X^n \!\mid\! \theta, M) \}, \\
\frac{d^2}{dt^2} F(t) &=& V_\theta^t \{ \log p(X^n \!\mid\! \theta, M) \}.
\end{eqnarray*}

Thus, we obtain the standard thermodynamic identity:
\begin{equation*}
\log L(M) = \int_0^1 \frac{d}{dt} F(t) dt = \int_0^1 E_\theta^t \{ \log p(X^n \!\mid\! \theta, M) \} dt. 
\end{equation*}

By the Cauchy-Schwarz inequality, we have $d^2 F(t)/dt^2 = V_\theta^t \{ \log p(X^n \!\mid\! \theta, M) \} > 0$. Therefore, $d F(t)/dt = E_\theta^t \{ \log p(X^n \!\mid\! \theta, M) \} $  is an increasing function. Hence, by the mean value theorem, there exists a unique temperature $t^* \in (0,1)$ such that
\begin{equation*}
\log L(M) = E_\theta^{t^*} \{ \log p(X^n \!\mid\! \theta, M) \}. 
\end{equation*}
Based on this fact, WBIC \citep{Watanabe2013} is defined as
\begin{equation*}
{\rm WBIC}  = E_\theta^{t_w} \{ \log p(X^n \!\mid\! \theta, M) \},
\end{equation*}
where $t_w = (\log n)^{-1}.$

The singular learning theory by \cite{Watanabe2009b} requires the following four assumptions.

\begin{assumption}
\label{assumption1}
The set of parameters $\Omega$ is a compact set in $R^d$ and can be defined by analytic functions $\pi_1,..., \pi_k$;
\begin{equation*}
\Omega = \{ \theta \in R^d : \pi_1(\theta)\ge 0,..., \pi_k(\theta)\ge 0 \}.
\end{equation*}
\end{assumption}

\begin{assumption}
\label{assumption2}
The prior distribution $\varphi(\theta)$ can be decomposed as the product of a non-negative analytic function $\varphi_1$ and a positive differentiable function $\varphi_2$;
\begin{equation*}
\varphi(\theta) = \varphi_1(\theta) \varphi_2(\theta).
\end{equation*}
\end{assumption}

\begin{assumption}
\label{assumption3}
Let $s \ge 6$ and
\begin{equation*}
L^s(q) = \left\{ f(x) : \Bigl(\int |f(x)|^s q(x)dx \Bigr)^{1/s} < \infty \right\}
\end{equation*}
be a Banach space. There exists an open set $\Omega' \supset \Omega$ such that for $\theta \in \Omega'$ the map $\theta \mapsto \log q(x)/p(x\mid\theta, M)$ is an $L^s(q)$-valued analytic function.
\end{assumption}

\begin{assumption}
\label{assumption4}
Let $\Omega_\epsilon$ be the set
\begin{equation*}
\Omega_\epsilon = \{ \theta \in \Omega : K(\theta) \leq \epsilon \},
\end{equation*}
where $K(\theta) = \int q(x) \log q(x)/p(x \!\mid\! \theta, M) dx$. There exists a pair of positive constants $(\epsilon, c)$ such that
\begin{equation*}
E \{ \log q(X)/p(X \!\mid\! \theta, M) \}  \ge c E \left[ \{ \log q(X)/p(X \!\mid\! \theta, M) \}^2 \right], \quad \forall \theta \in \Omega_\epsilon.
\end{equation*}
\end{assumption}

Under assumptions 1--4, \cite{Watanabe2009b} showed that
\begin{equation}
\log L(M) = \log p(X^n \!\mid\! \theta_0, M) - \lambda \log n + (m-1) \log \log n + O_p(1), \label{asymp_lml}
\end{equation}
where $ \theta_0$ is the parameter that minimizes the Kullback-Leibler divergence from a data-generating distribution to a statistical model, and $\lambda$ and $m$ are termed the real log canonical threshold and its multiplicity, respectively. The negative real log canonical threshold $(-\lambda)$ is defined as the largest pole of the zeta function $\zeta(z)$:
\begin{equation*}
\zeta(z) = \int_\Omega K(\theta)^z \varphi(\theta) d\theta, 
\end{equation*}
where $K(\theta) = \int q(x) \log q(x)/p(x \!\mid\! \theta, M) dx$ and $z$ is a complex variable. The multiplicity $m$ of the real log canonical threshold is defined as the order of the largest pole of the zeta function $\zeta(z)$.
Determining real log canonical thresholds and their multiplicities is generally challenging.

In addition, \cite{Watanabe2013} showed
\begin{equation*}
E( {\rm WBIC}) = E\{ \log p(X^n \!\mid\! \theta_0, M) \} - \lambda \log n + O(1).
\end{equation*}
In the next section, we identify the explicit constant order term.

\section{Constant order term of WBIC }

The Gibbs training loss $GL(t)$ is defined as
\begin{equation*}
GL(t) = - E_\theta^t \{ \log p(X^n \!\mid\! \theta, M) \}/n.
\end{equation*}
From the definition, we have
\begin{equation}
nGL(t_w) = - {\rm WBIC}. \label{gibbs_wbic}
\end{equation}
On the other hand, Theorems 6.8 and 6.10 of the book \citep{Watanabe2009b} lead to
\begin{equation}
E\{nGL(t) \} = -E\{ \log p(X^n \!\mid\! \theta_0, M) \} + \frac{\lambda}{t} -\nu(t) + o(1), \label{gibbs_exp}
\end{equation}
where $\nu(t)$ is called the singular fluctuation and is defined as
\begin{equation*}
\nu(t) = \lim_{n \to \infty} \frac{t}{2} E \left[ \sum_{i=1}^n  V_\theta^{t} \{ \log p({X}_i \!\mid\! {\theta}, M) \}  \right].
\end{equation*}
Equations (\ref{gibbs_wbic}) and (\ref{gibbs_exp}) lead to the following proposition:
\begin{proposition} \label{prop1}
Under assumptions 1--4, we have
\begin{equation}
E({\rm WBIC})  = E \left\{ \log p(X^n \!\mid\! \theta_0, M) \right\} - \lambda \log n + \nu(t_w) + o(1). \label{asymp_wbic}
\end{equation}
\end{proposition}
Since $\nu(t_w)$ is always positive by definition, equations (\ref{asymp_lml}) and (\ref{asymp_wbic}) cause WBIC to overestimate the leading two terms of the asymptotic expansion of the log marginal likelihood.

Let an estimator of the singular fluctuation $\hat{\nu}(t)$ be defined as
\begin{equation*}
\hat{\nu}(t) = \frac{t}{2} \left[ \sum_{i=1}^n  V_\theta^{t} \{ \log p({X}_i \!\mid\! {\theta}, M) \}  \right].
\end{equation*}
From Proposition \ref{prop1} and the definition of the singular fluctuation and its estimator, we obtain the following corollary.
\begin{corollary} \label{cor1}
Under assumptions 1--4, we have
\begin{equation*}
E\{ {\rm WBIC} - \hat{\nu}(t_w) \}  = E\left\{ \log p(X^n \!\mid\! \theta_0, M) \right\} - \lambda \log n + o(1). \label{asymp_wbic_nu}
\end{equation*}
\end{corollary}

\section{A simple example}

To demonstrate Corollary \ref{cor1}, we compute the explicit form of WBIC and $\hat{\nu}(t_w)$ for a simple model $M_N$ that was considered by \cite{Friel2008}, \cite{Friel2017}, and \cite{Gelman2013}. Let $x^n=\{x_i \mid i=1,...,n \}$ be independent and identically distributed observations, $x_i \sim N(\theta_0, 1)$, and the prior of $\theta$ is $N(m, v)$. Then, the posterior distribution is $N(m_t, v_t)$, where $m_t=(nt+1/v)^{-1}(nt\overline{x}+m/v), v_t=(nt+1/v)^{-1}$, and $\overline{x}=n^{-1}\sum_{i=1}^n x_i$. Therefore, a simple computation gives
\begin{eqnarray*}
{\rm WBIC} &=& -\frac{n}{2}\log 2\pi - \frac{1}{2}\left( \sum_{i=1}^n x_i^2 \right) + n \overline{x} m_{t_w} -\frac{n}{2}(v_{t_w} + m_{t_w}^2) \\
&=& -\frac{n}{2}\log 2\pi - \frac{1}{2}\left\{ \sum_{i=1}^n (x_i-\theta_0)^2 \right\}  - \frac{1}{2}\log n + \frac{n}{2}(\overline{x} -\theta_0)^2 +o_p(1) \\
&=& \log p(x^n \!\mid\! \theta_0, M_N) - \frac{1}{2}\log n + \frac{n}{2}(\overline{x} -\theta_0)^2 +o_p(1).
\end{eqnarray*}
In addition, for a large sample size $n$, the central limit theorem leads to
\begin{equation}
{\rm WBIC} = \log p(x^n \!\mid\! \theta_0, M_N) - \frac{1}{2}\log n + \frac{1}{2} +o_p(1). \label{wbic_simple_eg}
\end{equation}
On the other hand, a simple calculation leads to
\begin{equation*}
V_\theta^{t} \{ \log p({x}_i \!\mid\! {\theta}, M_N) \} = v_t(x_i - m_t)^2 + \frac{1}{2}v_t^2,
\end{equation*}
and from the definition of $\hat{\nu}(t)$,
\begin{equation*}
\hat{\nu}(t) = \frac{ntv - tv}{ntv+1}\frac{s_x^2}{2} + \frac{ntv}{2(ntv+1)^3}(m-\overline{x})^2 + \frac{tnv^2}{4(ntv+1)^2},
\end{equation*}
where $s_x^2=(n-1)^{-1}\sum_{i=1}^n (x_i-\overline{x})^2$.
Therefore, 
\begin{equation}
E\{ \hat{\nu}(t) \} = \frac{1}{2} + o(1), \label{nu_simple_eg}
\end{equation}
for any $t$. Equations (\ref{wbic_simple_eg}) and (\ref{nu_simple_eg}) lead to
\begin{equation*}
E\{ {\rm WBIC} - \hat{\nu}(t_w) \} = E\{ \log p(x^n \!\mid\! \theta_0, M_N) \} - \frac{1}{2}\log n +o(1).
\end{equation*}

\section{Numerical evaluation}
\subsection{Linear regression model}

The radiata pine dataset ($n=42$) was used in the book by \cite{Williams1959}, and $y_i$ denotes the maximum compression strength parallel to the grain, $x_i$ the density, and $z_i$ the resin-adjusted density. \cite{Friel2012} and \cite{Friel2017} considered the two non-nested linear regression models:
\begin{eqnarray*}
M_1: y_i &=& \alpha + \beta(x_i-\overline{x})+\epsilon_i, \quad \epsilon_i \sim N(0, \tau^{-1}), \\
M_2: y_i &=& \gamma + \delta(z_i-\overline{z})+\eta_i, \quad \eta_i \sim N(0, \kappa^{-1}),
\end{eqnarray*}
where $\overline{x}=n^{-1}\sum_{i=1}^n x_i$ and $\overline{z}=n^{-1}\sum_{i=1}^n z_i$.
They supposed the priors of $(\alpha, \beta)$ and $(\gamma, \delta)$ had mean $(3000,185)$ with precision $\tau Q$ and $\kappa Q$ respectively, where $Q$ is the diagonal matrix such that $Q_{(11)}=0.06, Q_{(22)}=6$. A gamma prior with shape $a=6$ and rate $b=600^2$ was chosen for $\tau$ and $\kappa$.

The exact evaluation of the log marginal likelihood was derived by \cite{Friel2012} and \cite{Friel2017}:
\begin{eqnarray*}
\log L(M_1) &=& -\frac{n}{2}\log \pi +\frac{a}{2}\log b + \log \frac{\Gamma\{(n+a)/2 \}}{\Gamma(a/2) } \nonumber \\
&&+ \frac{1}{2} \log \frac{\det (Q)}{\det (M)} - \frac{n+a}{2} \log(y^T R y + b), 
\end{eqnarray*}
where $y=(y_1,...,y_n)^T, M=X^TX + Q$, and $R=I- XM^{-1}X^T$ with $X$ the $n \times 2$ matrix such that $X_{(i1)}=1, X_{(i2)}=x_i$ and $I$ the $2 \times 2$ identity matrix. Obviously, $\log L(M_2)$ has the same expression.

We conducted 1000 independent computations of the Hamiltonian Monte Carlo method, implemented using the R package {\it RStan} \citep{RStan}, to obtain the posteriors for computing WBIC and WBIC $-\hat{\nu}(t_w)$. 

\begin{table}
\begin{center}
\caption{Radiata pine dataset ($n=42$). Comparison of the evaluations of the log marginal likelihood for linear regression models. Figures in parentheses give the standard deviations}
\begin{tabular}{ccccc}
\\ \hline
Methods & $\log L(M_{1})$ &  & $\log L(M_{2})$ &  \\
              & mean & s.d. & mean & s.d. \\ \hline
Exact Evaluation & -310.128 & - & -301.704 & - \\
WBIC & -308.091 & (0.0263) & -299.326 & (0.0272) \\
WBIC $-\hat{\nu}(t_w)$ & -310.100 & (0.0387) & -300.833 & (0.0365) \\ \hline
\end{tabular}
\label{tab1}
\end{center}
\end{table}

Table \ref{tab1} shows the results of the estimates of WBIC and WBIC $-\hat{\nu}(t_w)$.
Comparing WBIC and WBIC $-\hat{\nu}(t_w)$, the adjustment by the estimate of the singular fluctuation reduces the bias for both models.

\subsection{Normal mixture model}

In this section, we consider the following mixture model with two normal distributions:
\begin{equation*}
M_M: \alpha N(\mu_1, 1) + (1-\alpha) N(\mu_2, 1).
\end{equation*}
When the data-generating distribution is $N(0, 1)$, \cite{Aoyagi2010a} showed that the real log canonical threshold $\lambda$ is 3/4 and its multiplicity $m$ is 1.
Therefore, the log marginal likelihood $\log L(M_{M})$ is
\begin{equation*}
\log L(M_{M}) = \log p(X^n \!\mid\! \theta_0, M_M) - 3/4 \log n + O_p(1).
\end{equation*}

We conducted 1000 simulations to compute WBIC, WBIC $-\nu(t_w)$ and the Monte Carlo evaluation of $\log L(M_{M})$ for each sample size $n=50, 200$. 
We set the data-generating distribution $N(0, 1)$, and set the priors $\alpha \sim {\rm Unif}(0,1)$, $\mu_1, \mu_2 \sim N(0, 10)$. 
For Monte Carlo evaluation, we used standard Monte Carlo sampling with $10^7$ draws from the priors based on Neal's method \citep{Neal1999}.
We used the Hamiltonian Monte Carlo method, implemented with the R package {\it RStan} \citep{RStan}, to obtain the posteriors for computing WBIC and $\hat{\nu}(t_w)$.

\begin{table}
\begin{center}
\caption{Comparison of the approximations of the log marginal likelihood for Gaussian mixture models. Figures in parentheses give the standard deviations}
\begin{tabular}{ccccc}
\\ \hline
Methods & $n=50$ &  & $n=200$ & \\
              & mean & s.d. & mean & s.d.\\ \hline
Monte Carlo Evaluation & -74.32 & (4.56) & -288.84 & (9.61) \\ 
WBIC & -73.36 & (4.59) & -287.81 & (9.59) \\
WBIC $-\hat{\nu}(t_w)$ & -73.90 & (4.73) & -288.40 & (9.68) \\ \hline
\end{tabular}
\label{tab2}
\end{center}
\end{table}

Table \ref{tab2} shows the results of the estimates of WBIC, WBIC $-\hat{\nu}(t_w)$ and the Monte Carlo evaluation.
Comparing WBIC and WBIC $-\hat{\nu}(t_w)$, the values of the adjusted WBIC are closer to those of the Monte Carlo evaluation than those of WBIC.
The standard deviations of WBIC $-\hat{\nu}(t_w)$ are slightly larger than those of WBIC. 

\section{Discussion}

This paper identified the overestimating constant order term of WBIC, which is the singular fluctuation $\nu(t_w)$ with the temperature $t_w=(\log n)^{-1}$. The adjustment of WBIC by the estimator of singular fluctuation gives an asymptotically unbiased estimator for the leading two terms of the asymptotic expansion of the log marginal likelihood. Further work remains to be done regarding the higher asymptotic terms of the log marginal likelihood, including the term $(m-1) \log \log n$ and the $O_p(1)$ term in equation (\ref{asymp_lml}). Another future task is to construct an unbiased estimator of the singular fluctuation, which will reduce the bias of adjusting WBIC by the estimator of singular fluctuation when using a small sample size.

\section*{Acknowledgement}
The author was funded by the Japan Agency for Medical Research and Development.

\bibliographystyle{chicago}

\bibliography{WBICa.bib}
\end{document}